# Electron Charge Sensor with Hole Current Operating at Cryogenic Temperature

Digh Hisamoto[1], Noriyuki Lee[1]*, Ryuta Tsuchiya[1], Toshiyuki Mine[1], Takeru Utsugi[1], Shinichi Saito[1], and Hiroyuki Mizuno[1]

[1]*Research & Development Group, Hitachi, Ltd., Kokubunji, Tokyo 185-8601, Japan*

E-mail: dai.hisamoto.pd@hitachi.com

When SOI-PMOS functions like a capacitor-less 1T-DRAM cell, it is possible for the number of electrons to be sensed at cryogenic temperatures (5K). We developed a structure that combines SOI-NMOS and SOI-PMOS with multiple gates to form a silicon quantum-dot array. In this structure, a variable number of electrons is injected into the SOI-PMOS body transporting them by means of the bucket-brigade operation of SOI-NMOS connected in series. The channel-hole current was changed by the injected electrons due to the body bias effect in SOI-PMOS, and the change appeared to be step-like, suggesting a dependence on the elementary charge.

The scalability and long coherence time of Si quantum dot (QD)-based qubits make them promising candidates for large-scale quantum computing [1]–[6]. While electron spin states are used in computations, electron charges are exploited for measurement operations in quantum computing by spin-to-charge conversion [3]–[10]. Therefore, one of the keys to achieving Si quantum computers is the development of a sensor that can reliably read the number of electrons. To date, research has mainly focused on measurement using single-electron transistors (SETs) [2]–[9], or quantum point contacts (QPCs) [10] as charge sensors. RF reflectometry using a resonator has also been proposed [11]–[13]. However, to perform high-precision measurements, these techniques require the use of new devices such as SETs, QPCs, or resonators, which are not currently utilized in general LSIs.

In this paper, we propose an electronic charge sensor that can be implemented with digital LSI device technology because it uses only Silicon-on-Insulator p-type MOSFET (SOI-PMOS). When SOI-PMOS is executed in the same way as a capacitor-less one-transistor Dynamic Random Access Memory (1-T DRAM), electron charges can be converted into hole currents that amplify the signals, thus enabling highly sensitive electron-charge sensing [14]–[20]. This is what a Si quantum computer using electron spins needs to read out. In contrast to room temperature operation, electrons injected into the body of the PMOS at cryogenic temperatures have a much longer lifetime, resulting in a quasi-static current output [18]. Our idea is to combine this PMOS sensor with previously reported techniques for stable charge confinement and transport in MOS structures that take advantage of the cryogenic properties of low thermal fluctuations [21]–[29], thereby enabling the use of standard LSI technology for the measurement of Si quantum computers.

With the conventional Si quantum dot 1-D array as a basis [3], [8], [28], we utilized an SOI wafer, and formed a series-connected small n-type MOSFET (NMOS) that act as qubits and barrier gates when a single electron is stored. We also replaced the SET sensor with PMOS one and formed a connecting gate between the quantum dot and the sensor. In this structure, a quantum dot adjacent to the sensor first holds a defined number of electrons through spin-to-charge conversion operation and then transports them to the body of the SOI-PMOS. Consequently, the change in channel current due to the PMOS body bias effect should allow the number of electrons to be determined. Here, we report the behavior of SOI-PMOS sensors at cryogenic temperatures, as spin-to-charge conversion operation and charge transport have already been sufficiently investigated [3]–[10], [21]–[26].

Figure 1 shows, the structure of the Si channels and poly-Si gates (separated by color) in a top-view scanning electron microscope (SEM) image of the devices. We used an SOI wafer with a 145-nm-thick buried oxide (BOX) layer, as in our previous work [29]–[31]. The Si channel is intrinsic, so both electrons and holes can flow through it. The Si channel (colored green in Fig. 1) has three terminals: p-type source (S), p-type drain (D), and n-type reservoir (R). Three layers of gates are formed above the Si channel: first gates (FGs, red), second gates (SGs, blue), and third gates (TGs, orange). SGs are formed by a self-aligned patterning process for FGs, and TGs are then formed by a self-aligned patterning process for SGs. The device parameters are listed in Table I. In the sensing path of Fig. 1, a hole current flows between the p-type source and drain under three gates labeled as TG1, SGS, and TG2, where SGS acts as a PMOS sensor, and TG1 and TG2, which are always set to "on", act as extensions of the source and drain electrodes. The other gates (FG0–3, SG1–3) next to the n-type reservoir in the electron injection path in Fig. 1 are utilized to transport the required number of electrons from the reservoir to SG3-dot. S and D act as the source and drain of the PMOS sensor, and R supplies electrons to the PMOS body as shown in the inset of Fig. 1. Hereinafter, to distinguish between the gate electrode and the SOI channel portion, the channel portion is denoted by "-dot". If SG2-dot and SG3-dot are used for spin-to-charge conversion operation, a limited number of electrons can be put into SG3-dot. Then, all the electrons accumulated in SG3-dot can be transferred to the PMOS body (SGS-dot) using FG3.

We performed all the electrical characterizations in a cryogenic probe station using a semiconductor parameter analyzer. To investigate the amplification properties of the PMOS sensor, instead of spin-to-charge conversion, we utilized a gating sequence to change the number of electrons injected into the SG3 dots from the reservoir. A cross-sectional view of the part for supplying electrons (A-A' in Fig. 1) is shown in Fig. 2 (a). To observe the electrical properties of FG and SG at each temperature, we fabricated devices with n-type drains and measured them (Fig. 2(b) and (c)). The threshold voltage ($V_{th}$) between FG and SG was different due to the different gate work function (see Table I). We also found that the $V_{th}$ and subthreshold swing (SS) of FG and SG exhibited similar characteristics to the previously reported temperature dependence [18], [30]. Considering these electrical characteristics at 5 K, we devised a sequence for supplying electrons to SG3-dot.

As shown in the first two steps in Fig. 2(d), the electrons are completely swept into the reservoir by applying a negative potential sequentially from the gate closest to the SGS as initialization. Then, the five FG and SG gates (FG0 to FG2) that create the potential barrier between the reservoir and SG3 are fixed, and the potential barrier height is changed by changing the reservoir voltage ($V_R$). This controls the number of electrons injected into SG3-dot (the third step in Fig. 2(d)). Raising the potential barrier (FG2) and returning $V_R$ to ground level terminates the injection operation and fixes the electron number of SG3-dot (the fourth step). Since $V_R$ returns to the ground level, its value during electron injection is denoted as $V_{R,inj}$. A smaller $V_{R,inj}$ lowers the potential barrier during electron injection, thus allowing more electrons to be injected into SG3-dot. Then, the confined electrons are shuttled from SG3-dot to the body of the PMOS (SGS-dot) [21], [22]. Consequently, the number of electrons confined inside the PMOS is controlled by the $V_{R,inj}$.

The operating principle of the PMOS sensor based on a capacitor-less 1T-DRAM is depicted in Fig. 3 (a) and (b), which are cross-sectional views and energy diagrams of the part that senses electrons (B-B' in Fig. 1). When electrons are injected into SGS-dot, the electron charges change the conduction band ($E_C$) and valence band ($E_V$) energy levels below the SGS. This change tends to make the barrier larger for electrons but lowers the barrier for holes, thus increasing the PMOS channel current. As a result, the number of electrons is converted into a static PMOS current, so a large signal amplification can be achieved. Note that as previously reported, carriers injected into the MOSFET body have long lifetimes at cryogenic temperatures [18]. As a preliminary experiment, we measured the time dependence of the PMOS output current after charge injection, and confirmed that the current change was within the noise floor level of the measurement system even 10 minutes after the injection. Since electrons are stably trapped in SGS-dots, they can be transferred back to SG3-dots after measurement if desired.

In the measurements, we first characterized the dependence of the PMOS on the $V_R$. Figure 3(c) and (d) show $I_S$-$V_{SGS}$ characteristics of the PMOS with $V_S$ = 0.2 V, $V_D$ = 0.0 V, and $V_{TG1}$ = $V_{TG2}$ = –2.3 V at room temperature (300 K) and cryogenic temperature (5 K). We changed the $V_R$ from 0.6 V to –0.2 V while turning on all the gates between the reservoir and the PMOS by applying 2.0 V to the FGs and 0.5 V to the SGs. At both temperatures, the PMOS Vth shift depended on $V_R$ due to body bias effects. Comparing the measurements at both temperatures, we found that Vth shifted in the negative direction and the subthreshold slope steepened at 5K. These results suggest that PMOS can work as a sensor even at 5 K.

Next, using the operation sequence described in Fig. 2(d), we measured the conversion characteristics from the amount of electrons of the PMOS sensor injected to the channel current. After electron confinement, we set all voltage conditions (including $V_R$) to the same value before measuring the hole current to mitigate the effect of coupling between the electrodes and observe the change in current with the amount of injected electrons. The $V_{R,inj}$ determines the number of electrons confined in SGS-dot. Therefore, the PMOS substrate potential is set by $V_{R,inj}$, not by $V_R$ when measuring current. Figure 4(a) shows the PMOS current $I_S$ characteristics measured by sweeping $V_{SGS}$ from –0.5 V to –1 V while changing $V_{R,inj}$ in increments of 0.1 V in the electron supply sequence at 5 K. The integration time per measurement point is 80 ms. We observed that $I_S$ increased as $V_{R,inj}$ decreased. Conversely, when $V_{R,inj}$ increased and exceeded 0.3 V, there were no electrons injected into the PMOS body, so $I_S$ showed a constant value. This can be confirmed from the overlapping $I_S$ waveforms of $V_{R,inj}$= 0.3 V and 0.4 V. If $V_{SGS}$ is sufficiently lower than $V_{FG3}$, the electrons held in the PMOS body will be ejected. In this measurement with SGS operation above –1.0 V, electrons appeared to be trapped inside the PMOS body due to the difference in Vth between FG3 and SGS.

To investigate the sensitivity of $I_S$ to $V_{R,inj}$ in detail, we measured $I_S$ while changing $V_{R,inj}$ from 0.3 V to 0.15 V in 10-mV steps. Measurements were repeated three times for each step. Figure 4(b) plots the value of $I_S$ at $V_{SGS}$ = –0.7 V obtained from the current value versus $V_R$,inj for each measurement. The increase in $I_S$ is not a continuous change, but shows discrete values in three levels, that is, base-level, 1st-level, and 2nd-level as shown in Fig.4(b). When $V_{R,inj}$ was 0.25 V or higher, $I_S$ exhibited a nearly constant minimum value (base-level) because no electrons are injected. When $V_{R,inj}$ was below 0.25 V, $I_S$ increased by taking one or two different levels for the same $V_{R,inj}$ and moving to higher levels.

To analyze this discrete increase in $I_S$, we estimated the conversion efficiency due to the PMOS body bias effect. A hole current flows on the top surface of the SOI and the confined electrons are distributed on the bottom surface, so the channel surface potential $\varphi_S$ is modeled by the series connection of the capacitance $C_{OX}$ between the gate and the channel and the capacitance $C_{SOI}$ between the channel and the bottom of the SOI (see Fig. 3 (a)) [15]. Other capacitive components, such as capacitance with the source, can be considered constant because they are measured under the same bias conditions, and are connected to $\varphi_S$ in parallel with $C_{OX}$ and $C_{SOI}$. Therefore, when considering the effect on $\varphi_S$ due to injected charge into the gate electrode or the SOI body, we can model without them. By using the capacitance ratio $C_{OX} / C_{SOI}$, the change in channel potential due to the charge on the gate electrode can be converted into the change due to the charge on the SOI body. $C_{OX}$ is estimated to be $1.6×10^{-17}$ F from the area (80 nm × 90 nm) and the gate oxide thickness (15 nm) and the 50 nm SOI thickness gives a $C_{OX} / C_{SOI}$ of 1.10. We can therefore use this capacitor model to estimate the potential change due to electron charges in the body and determine the change in channel current. We obtained the voltage-to-current conversion efficiency of measured IV characteristics. At the measurement condition $V_{SGS} = –0.7$ V, the efficiency ($\Delta I/\Delta V_{SGS}$) is estimated to be $2.6×10^{-7}$ A/V for the base-level, and $3.5×10^{-7}$ A/V for the 1st-level. Therefore, if an elementary charge of $1.6×10^{-19}$ C is placed on the PMOS body under this operating condition, the current in the PMOS will increase by 2.8 nA from the base-level state and 3.7 nA from the first-level state. As shown in Fig. 4(b), these values are in good agreement with the discrete variation of $I_S$, suggesting that the discrete variation of $I_S$ was caused by the electron charge in the body of the PMOS.

Regarding the control step of the amount of injected charge in the experimental sequence, multiple levels of $I_S$ were observed for the same $V_{R,inj}$, suggesting that the electron number controllability of the sequence is still insufficient. On the other hand, the observation of step-like changes suggests that a precise manipulation of the number of electrons was performed. In the experimental sequence, one large potential barrier was schematically drawn in the third step of Fig. 2(d). However, in reality, multiple potential barrier peaks exist like a mountain range between the reservoir and the SG3dots due to the difference in gate work function between FG and SG. Therefore, since the $V_{R,inj}$ changes in fine increments (10 mV), it seems that the dominant potential barrier acted like a tunnel barrier, delicately manipulating the injected electrons one by one. We also presumed that when electrons are transported in this order, some observation points are out of step because the parasitic holes are not swept out sufficiently, thus leaving charges in unwanted places. In the future, it will be necessary to utilize an established single-electron manipulation sequence (e.g., pump operation [21]–[26]) to transport single electrons into the PMOS sensor and verify its operation.

Finally, we touch on whether the PMOS can be controlled by the peripheral circuits of an LSI. In a conventional general-purpose memory, the output current from a memory cell (equivalent to the PMOS sensor here) drives a gate in a post-stage amplifier circuit to output a differential voltage of 100 mV. The PMOS current of 2.8 nA can charge the gate capacitance of 0.25 fF (L = 200 nm and W = 180 nm) within 9 ns. Therefore, using the proposed PMOS sensor in a quantum computer, it should be possible to perform standard 100-MHz read operations using conventional peripheral circuits.

In conclusion, we demonstrated that a stored electron charge can be sensed using the body bias effect in an SOI-PMOS at cryogenic temperatures. This sensor can be implemented with conventional LSI technology and has high sensitivity at the elementary charge level, so the combination of spin-to-charge conversion and this sensor has the potential to perform the required measurements in Si quantum computing operations. Although we used a PMOS in this work, it is also possible to observe holes in an NMOS for a quantum computer with hole spin qubits [32]–[34]. We believe that the ability to simultaneously utilize electrons and holes is the greatest feature of Si technology and will lead to new developments for Si quantum computers.

**Acknowledgments**

This work was supported by JST Moonshot R&D (Grant Number JPMJMS2065).

## References

1) D. Loss and D. P. DiVincenzo, Phys. Rev. A **57**, 120 (1997).
2) M. Veldhorst, C. H. Yang, J. C. C. Hwang, W. Huang, J. P. Dehollain, J. T. Muhonen, S. Simmons, A. Laucht, F. E. Hudson, K. M. Itoh, A. Morello, and A. S. Dzurak, Nature **526**, 410 (2015).
3) C. H. Yang, W. H. Lim, F. A. Zwanenburg, and A. S. Dzurak, AIP Advances **1**, 042111 (2011).
4) A. Morello, J. J. Pla, F. A. Zwanenburg, K. W. Chan, K. Y. Tan, H. Huebl, M. Möttönen, C. D. Nugroho, C. Yang, J. A. van Donkelaar, A. D. C. Alves, D. N. Jamieson, C. C. Escott, L. C. L. Hollenberg, R. G. Clark, and A. S. Dzurak, Nature **467**, 687 (2010).
5) M. Veldhorst, J. C. C. Hwang, C. H. Yang, A. W. Leenstra, B. de Ronde, J. P. Dehollain, J. T. Muhonen, F. E. Hudson, K. M. Itoh, A. Morello, and A. S. Dzurak, Nat. Nanotechnol. **9**, 981 (2014).
6) K. Takeda, J. Kamioka, T. Otsuka, J. Yoneda, T. Nakajima, M. R. Delbecq, S. Amaha, G. Allison, T. Kodera, S. Oda, and S. Tarucha, Science Advances **2**, e1600694 (2016).
7) J. Yoneda, K. Takeda, T. Otsuka, T. Nakajima, M. R. Delbecq, G. Allison, T. Honda, T. Kodera, S. Oda, Y. Hoshi, N. Usami, K. M. Itoh, and S. Tarucha, Nat. Nanotechnol. **13**, 102 (2018).
8) A. E. Seedhouse, T. Tanttu, R. C. C. Leon, R. Zhao, K. Y. Tan, B. Hensen, F. E. Hudson, K. M. Itoh, J. Yoneda, C. H. Yang, A. Morello, A. Laucht, S. N. Coppersmith, A. Saraiva, and A. S. Dzurak, PRX Quantum **2**, 010303 (2021).
9) A. C. Johnson, J. R. Petta, C. M. Marcus, M. P. Hanson, and A. C. Gossard, Phys. Rev. B **72**, 165308 (2005).
10) J. M. Elzerman, R. Hanson, L. H. Willems van Beveren, B. Witkamp, L. M. K. Vandersypen, and L. P. Kouwenhoven, Nature **430**, 431 (2004).
11) M. F. Gonzalez-Zalba, S. N. Shevchenko, S. Barraud, J. R. Johansson, A. J. Ferguson, F. Nori, and A. C. Betz, Nano Lett. **16**, 1614 (2016).
12) M. F. Gonzalez-Zalba, S. Barraud, A. J. Ferguson, and A. C. Betz, Nat. Commun. **6**, 6084 (2015).
13) S. Schaal, A. Rossi, V. N. Ciriano-Tejel, T. Yang, S. Barraud, J. J. L. Morton, and M. F. Gonzalez-Zalba, Nat. Electron. **2**, 236 (2019).
14) P. F. Lu, C. T. Chuang, J. Ji, L. F. Wagner, C. M. Hsieh, J. B. Kuang, L. L. Hsu, M. M. Pelella, S. Chu, and C. J. Anderson, IEEE J. Solid-State Circuits **32**, 1241 (1997).
15) A. Wei, M. Sherony, and D. A. Antoniadis, IEEE Trans. Electron Devices **45**, 430 (1998).

16) M. Bawedin, S. Cristoloveanu, and D. Flandre, Solid-State Electronics **51**, 1252 (2007).

17) I. Ban, U. E. Avci, D. L. Kencke, and P. L.D. Chang, Symp. VLSI Tech., 2008, p. 92.

18) W. Chakraborty, R. Saligram, A. Gupta, M. S. Jose, K. A. Aabrar, S. Dutta, A. Khanna, A. Raychowdhury, and S. Datta, IEDM Tech. Dig., 2021, p. 853.

19) E. Yoshida, T. Tanaka, IEEE Trans. Electron Devices **53**, 692 (2006).

20) M. Bawedin, S. Cristoloveanu, and D. Flande, IEEE Electron Device Lett., **29**, 795 (2008).

21) A. Fujiwara and Y. Takahashi, Nature **410**, 560 (2001).

22) T. A. Baart, M. Shafiei, T. Fujita, C. Reichl, W. Wegscheider, and L. M. K. Vandersypen, Nat. Nanotechnol. **11**, 330 (2016).

23) K. Nishiguchi, A. Fujiwara, Y. Ono, H. Inokawa, and Y. Takahashi, Appl. Phys. Lett. **88**, 183101 (2006).

24) G. Yamahata, S. P. Giblin, M. Kataoka, T. Karasawa, and A. Fujiwara, Scientific Reports **7**, 45137 (2017).

25) G. Yamahata, S. Ryu, N. Johnson, H.-S. Sim, A.Fujiwara, and M. Kataoka, Nat. Nanotechnol. **14**, 1019 (2019).

26) B. Kaestner, V. Kashcheyevs, S. Amakawa, M. D. Blumenthal, L. Li, T. J. B. M. Janssen, G. Hein, K. Pierz, T. Weimann, U. Siegner, and H. W. Schumacher, Phys. Rev. B **77**, 153301 (2008).

27) T. Nakajima, A. Noiri, J. Yoneda, M. R. Delbecq, P. Stano, T. Otsuka, K. Takeda, S. Amaha, G. Allison, K. Kawasaki, A. Ludwig, A. D. Wieck, D. Loss, and S. Tarucha, Nat. Nanotechnol. **14**, 555 (2019).

28) J. Yoneda, W. Huang, M. Feng, C. H. Yang, K. W. Chan, T. Tanttu, W. Gilbert, R. C. C. Leon, F. E. Hudson, K. M. Itoh, A. Morello, S. D. Bartlett, A. Laucht, A. Saraiva, and A. S. Dzurak, Nat. commun. **12**, 4114 (2021).

29) T. Utsugi, N. Lee, R. Tsuchiya, T. Mine, G. Shinkai, I. Yanagi, Y. Kanno, T. Sekiguchi, S. Akiyama, T. Norimatsu, Y. Wachi, R. Mizokuchi, J. Yoneda, T. Kodera, S. Saito, D. Hisamoto, and H. Mizuno, Ext. Abstr. Solid State Devices and Materials, 2022, p. 513.

30) N. Lee, R. Tsuchiya, G. Shinkai, Y. Kanno, T. Mine, T. Takahama, R. Mizokuchi, T. Kodera, D. Hisamoto, and H. Mizuno, Appl. Phys. Lett. **116**, 162106 (2020).

31) N. Lee, R. Tsuchiya, Y. Kanno, T. Mine, Y. Sasago, G. Shinkai, R. Mizokuchi, J. Yoneda, T. Kodera, C. Yoshimura, S. Saito, D. Hisamoto, and H. Mizuno, Jpn. J. Appl. Phys. **61**, SC1040 (2022).

32) N. Piot, B. Brun, V. Schmitt, S. Zihlmann, V. P. Michal, A. Apra, J. C. Abadillo-Uriel, X.

Jehl, B. Bertrand, H. Niebojewski, L. Hutin, M. Vinet, M. Urdampilleta, T. Meunier, Y.-M. Niquet, R. Maurand, and S. De Franceschi, arXiv:2201.08637.

33) F. N. M. Froning, L. C. Camenzind, O. A. H. van der Molen, A. Li, E. P. A. M. Bakkers, D. M. Zumbühl, and F. R. Braakman, Nat. Nanotechnol. **16**, 308–312 (2021).

34) L. C. Camenzind, Simon Geyer, A. Fuhrer, R. J. Warburton, D. M. Zumbuhl, and V. Kuhlmann, Nat. Electron., 5, 178-183 (2022).

**Figure Captions**

**Fig. 1.** Top-view SEM image with colored schematics of device and circuit diagram (inset). T-shaped SOI active region (green) has three terminals: p-type source (S), p-type drain (D), and n-type reservoir (R). Three-layered poly-Si gates (red, blue, orange) are arranged. SG3 defines quantum dot and SGS constitutes PMOS sensor.

**Fig. 2.** (a) Cross-sectional view of TEM image for the electron-injection path. (b), (c) FG3 and SG3-NMOS characteristics at various temperatures. $V_D = 0$ V, $V_R = 0.2$ V and $V_S$ is open. (d) Sequence for transporting electrons from reservoir to PMOS-dot. The right side shows the transport of carriers at each step using a schematic diagram of the surface potential distribution.

**Fig. 3.** (a) Cross-sectional view of TEM image for the sensing path. (b) Operating principle of PMOS sensor. The top is before electron injection and the bottom is after electron injection. PMOS $I_s$-$V_{SGS}$ characteristics with various $V_R$ at (c) 300 K and (d) 5 K.

**Fig. 4.** (a) PMOS $I_s$-$V_{SGS}$ characteristics at 5 K with $V_{R,inj}$ as a parameter. $V_{R,inj}$ controls electrons supplied to PMOS body (SGS-dot) by the sequence. (b) Current flowing through PMOS as function of $V_{R,inj}$ with $V_{SGS} = -0.7$ V at 5 K.

**Table I.** Device parameters.

|  | First gate (FG) | Second gate (SG) | Third gate (TG) |
|---|---|---|---|
| $T_{OX}$ | 5 nm | 15 nm | 15 nm |
| $W$ | 50 nm | 50 nm | 50 nm |
| $L$ | 80 nm | 50 nm | 200 nm |
| $T_{SOI}$ | 50 nm | 50 nm | 50 nm |
| $T_{BOX}$ | 145 nm | 145 nm | 145 nm |
| Gate | B-doped (P+) | P-doped (N+) | P-doped (N+) |
| Channel | non-doped | non-doped | non-doped |

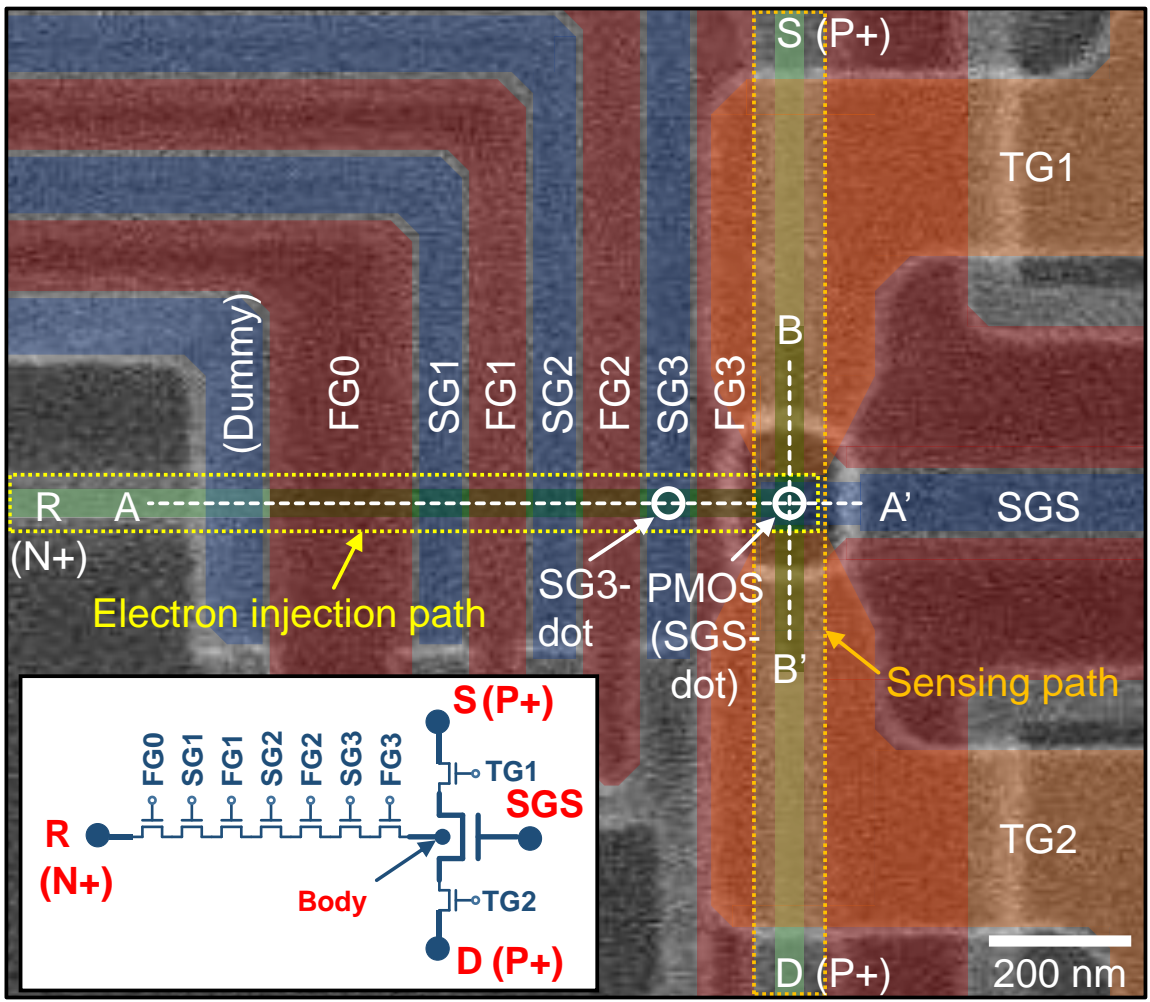

Fig.1.

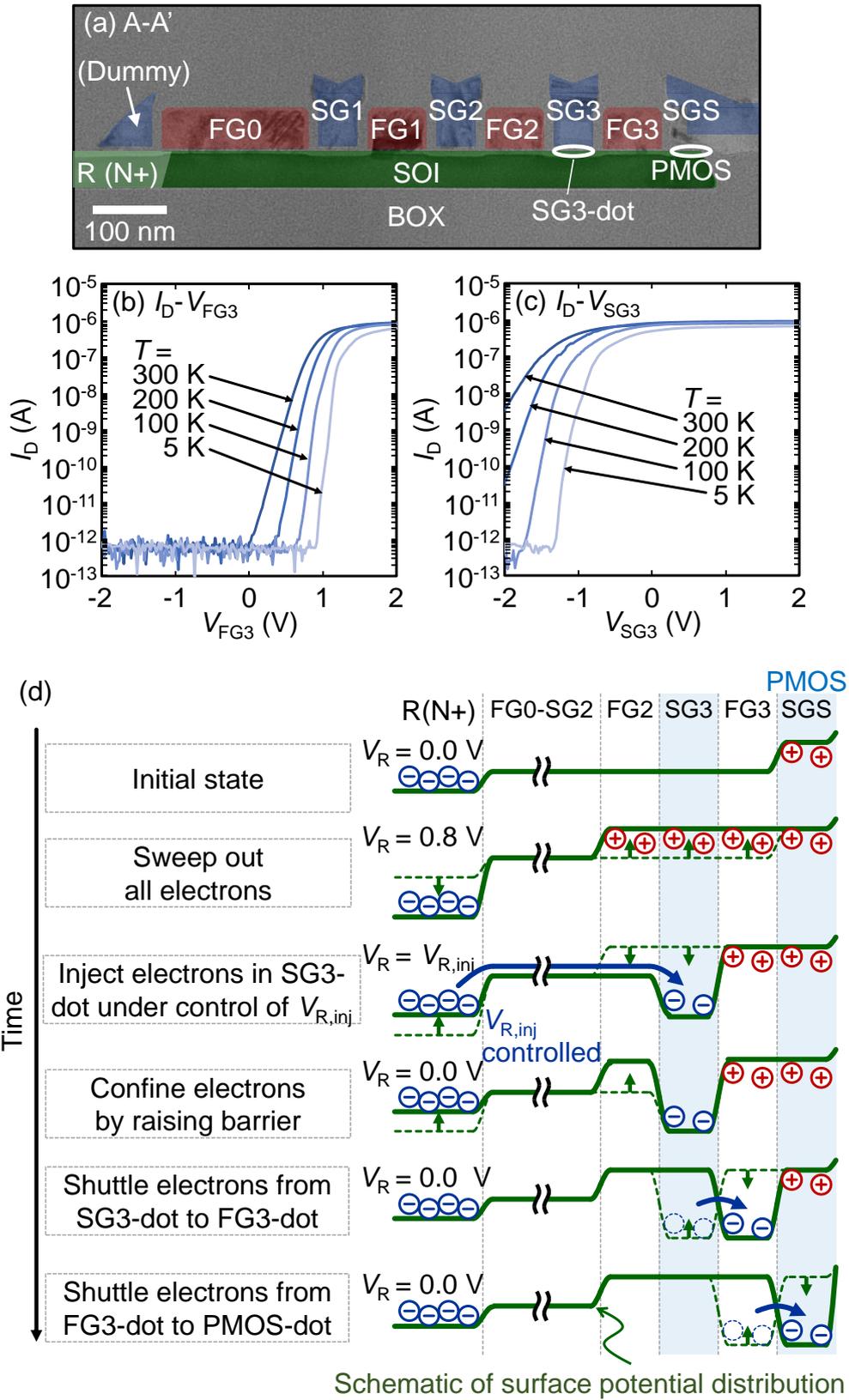

Fig. 2.

Fig. 3.

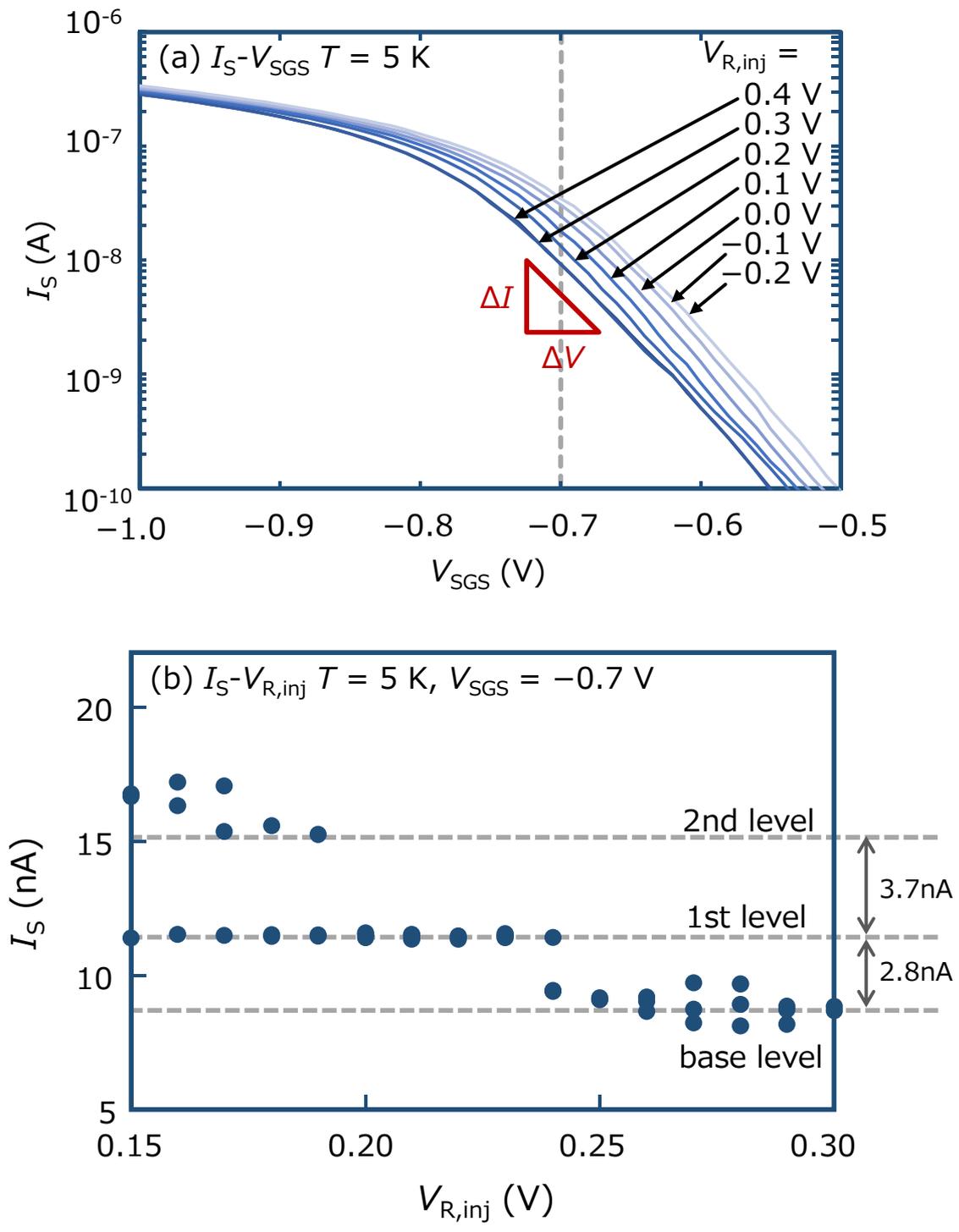

Fig. 4.